\documentclass{article}
\usepackage{mdframed}
\usepackage[most]{tcolorbox}
\usepackage{todonotes}
\usepackage{PRIMEarxiv}
\usepackage[utf8]{inputenc} 
\usepackage[T1]{fontenc}    
\usepackage{hyperref}  
\usepackage{amssymb,amsmath,amsthm}
\usepackage{xspace}
\usepackage{mathtools}
\usepackage{graphicx,xcolor}
\usepackage{url}            
\usepackage{booktabs}       
\usepackage{amsfonts}       
\usepackage{nicefrac}       
\usepackage{microtype}      
\usepackage{lipsum}
\usepackage{fancyhdr}       
\usepackage{graphicx} 
\usepackage[ruled,linesnumbered]{algorithm2e}
\usepackage[utf8]{inputenc}
\graphicspath{{media/}}     

\newcommand{\np}{\textup{\textsf{NP}}\xspace}

\newcommand{\mcs}{\textup{\textsc{MCS}}\xspace}
\newcommand{\ptas}{\textup{\textsc{PTAS}}\xspace}

\newcommand{\cs}{\textup{\textsc{CS}}\xspace}
\newcommand{\scs}{\textup{\textsc{SCS}}\xspace}
\newcommand{\mscs}{\textup{\textsc{MSCS}}\xspace}
\newcommand{\mcss}{\textup{\textsc{MCSS}}\xspace}
\newcommand{\css}{\textup{\textsc{CSS}}\xspace}

\newcommand{\NN}{\mathrm{NN}\xspace}

\newcommand{\apx}{\textup{\textsf{APX}}\xspace}

\newcommand{\patha}{\mathrm{path}\xspace}
\newcommand{\paath}{\mathrm{path}\xspace}
\newcommand{\dist}{\mathrm{d}\xspace}

\newcommand{\neigh}{\mathrm{N}\xspace}

\newcommand{\tree}{\mathrm{T}\xspace}

\newcommand{\xp}{{\sf XP}\xspace}
\newcommand{\fpt}{{\sf FPT}\xspace}

\usepackage{lineno}

\newtheorem{theorem}{Theorem}[section]

\newtheorem{lemma}[theorem]{Lemma}

\newtheorem{observation}{Observation}[theorem]
\title{Minimum Strict Consistent Subset in Paths, Spiders, Combs and Trees
}
\author{
  Bubai Manna\\
  Department of Mathematics, Indian Institute of Technology Kharagpur, India. \\
  \texttt{bubaimanna11@gmail.com}
}
\begin{document}
\maketitle
\begin{abstract}
 Let $G$ be a simple connected graph with vertex set $V(G)$ and edge set $E(G$. Each vertex of $V(G)$ is colored by a color from the set of colors $\{c_1, c_2,\dots, c_{\alpha}\}$. We take a subset $S$ of $V(G)$, such that for every vertex $v$ in $V(G)\setminus S$, at least one vertex of the same color is present in its set of nearest neighbors in $S$. We refer to such an $S$ as a consistent subset (\cs). The Minimum Consistent Subset (\mcs) problem is the computation of a consistent subset of the minimum cardinality. It is established that \mcs is NP-complete for general graphs, including planar graphs. The strict consistent subset is a variant of consistent subset problems. We take a subset $S^{\prime}$ of $V(G)$, such that for every vertex $v$ in $V(G)\setminus S^{\prime}$, all the vertices in its set of nearest neighbors in $S^{\prime}$ have the same color as that of $v$. We refer to such an $S^{\prime}$ as a strict consistent subset (\scs). The Minimum Strict Consistent Subset (\mscs) problem is the computation of a strict consistent subset of the minimum cardinality.

\smallskip

We demonstrate that \mscs is \np-hard for general graphs using a reduction from dominating set problems. We construct a 2-approximation algorithm and a polynomial-time algorithm in trees. Lastly, we conclude the faster polynomial-time algorithms in paths, spiders, and combs.

\vspace{-0.1in}
\end{abstract}


\section{Introduction}\label{intro1}

Many supervised learning approaches use a colored training dataset $T$ in a metric space $(X, d)$ as input, with each element $t \in T$ assigned a color $c_i$ from the set of colors $C=\{c_1, c_2, \dots, c_{\alpha}\}$. The purpose is to preprocess $T$ in order to meet specific optimality requirements by immediately assigning color to any uncolored element in $X$. The closest neighbor rule, which assigns a color to each uncolored element $x$ based on the colors of its $k$ closest neighbors in the training dataset $T$ (where $k$ is a fixed integer), is a popular optimality criterion. The efficiency of such an algorithm is determined by the size of the training dataset. As a result, it is necessary to reduce the cardinality of the training set while retaining the distance data. Hart \cite{Hart} introduced this concept in $1968$ with the minimum consistent subset (in short\mcs) problem. Given a colored training dataset $T$, the goal is to find a subset $S \subseteq T$ of the minimum cardinality such that the color of every point $t \in T$ is the same as one of its nearest neighbors in $S$. Over $2800$ citations to \cite{Hart} on Google Scholar demonstrate that the \mcs problem has found several uses since its inception. \cite{Wilfong} demonstrated that the \mcs problem for points in $\Re^2$ under the Euclidean norm is \np-complete if at least three colors color the input points. Furthermore, it is \np-complete even with two colors \cite{Bodhayan}. Recently, it was established that the \mcs problem is W[1]-hard when parameterized by the output size \cite{Chitnis}.\cite{Wilfong} and \cite{Biniaz} offer some other algorithms for the \mcs problem in $\Re^2$.

\subsection{Notations and Definitions}\label{intro} 

This paper explores the minimum consistent subset problem when $(\mathcal{X},d)$ is a graph metric. Most graph theory symbols and notations are standard and come from \cite{diestel2012graph}. For every graph $G$, we refer to the set of vertices as $V(G)$, and the set of edges as $E(G)$. Consider a graph $G$ with a vertex coloring function $C:V(G) \rightarrow \{c_1, c_2, \dots, c_{\alpha}\}$. For any subset $U$ of $V(G)$, let $C(U)$ represent the set of colors of the vertices in $U$. Formally, $C(U)=\{C(u):u\in U\}$. For any two vertices $u,v \in V(G)$, we use $\dist(u,v)$ to signify the shortest path distance (number of edges) between $u$ and $v$ in $G$. $\dist(v,U)=\min_{u\in U}\dist(v,u)$ for a vertex $v\in V(G)$. 

\smallskip

For any graph $G$ and any vertex $v\in V(G)$, let $\neigh(v)=\{u\mbox{ }|\mbox{ }u\in V(G), (u,v)\in E(G)\}$ indicate the set of neighbors of $v$ and $\neigh[v]=\{v\}\cup \neigh(v)$. For $U\subset V$, $\neigh [U]= \cup_{u\in U}\neigh [u]$. The distance between two subgraphs $G_1$ and $G_2$ in $G$ is represented as $\dist(G_1,G_2)=\min \{\dist(v_1,v_2)\mbox{ }|\mbox{ }v_1\in V(G_1),v_2\in V(G_2)\}$. Therefore, $\neigh ^{r}[v]=\{u\in V \mbox{ }|\mbox{ }\dist (u,v)\leq r \}$. The nearest neighbor of $v$ in the set $U$ is indicated as $\NN(v,U)$, which is formally defined as $\{u\in U \mbox{ }| \mbox{ }\dist(v,u)=\dist(v,U)\}$. Also, for $X, U\subseteq V$, $\NN(X,U)=\cup_{v\in X}\{u\in U \mbox{ }| \mbox{ }\dist(v,u)=\dist(v,U)\}$. The shortest path in $G$ is $\patha (v,u)$, which connects vertices $v$ and $u$. $\patha (v,U)=\{\patha (v,u) \mbox{ }|\mbox{ }u\in \NN(v,U)\}$. Without loss of generality, we shall use $[n]$ to denote the set of integers $\{1, \ldots, n\}$. A subset $X$ is said to be a \emph{consistent subset} (in short \cs) for $(G,C)$ if every $v\in V(G)$, $C(v)\in C(\NN(v,X))$. A subset $S$ is said to be a \emph{strict consistent subset} (in short \scs) for $(G,C)$ if every $v\in V(G)$, $C(v)= C(\NN(v,S))$. \mcs is a \cs with minimum cardinality. Similarly, a minimum strict consistent subset (in short \mscs) refers to a strict consistent subset with minimum cardinality. The strict consistent subset problem in graphs is defined as follows:

\begin{tcolorbox}[enhanced,title={\color{black} \sc{\mbox{    }\mscs on Graphs}}, colback=white, boxrule=0.4pt, attach boxed title to top center={xshift=-4cm, yshift*=-2mm}, boxed title style={size=small,frame hidden,colback=white}]
		
    \textbf{Input:} A graph $G$, a coloring function $C:V(G)\rightarrow \{c_1, c_2, \dots, c_{\alpha}\}$, and an integer $\ell$.\\
    \textbf{Question:} Does there exist a strict consistent subset of size $\le \ell$ for $(G,C)$?
\end{tcolorbox}
 
\smallskip

\begin{figure}[t]
\includegraphics[width=12cm]{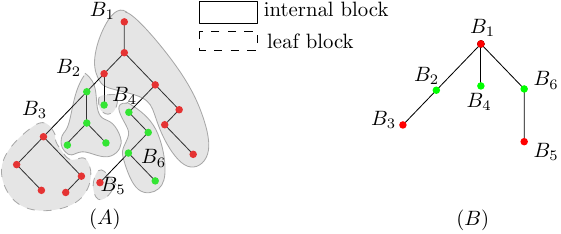}
\centering
\caption{(A) Blocks are $B_1, \dots B_6$. (B) The corresponding block tree.}\label{1b}
\end{figure}

\begin{figure}[t]
\includegraphics[width=12cm]{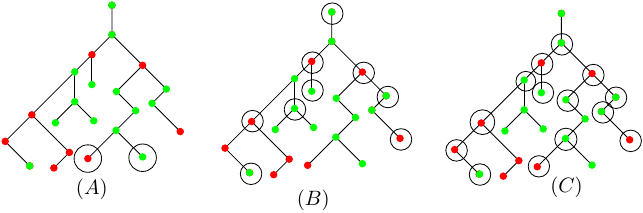}
\centering
\caption{(A), (B) and (C) are the examples of \mcs, \mcss and \mscs for a two-colored unweighted tree, respectively.}\label{1a}
\end{figure}

For a \scs $S$ and a vertex $v\in V(G)\setminus S$, $v$ is said to be \emph{covered} by a vertex $u\in \NN(v,S)$ if $C(u)=C(v)=C(\NN(v,S))$. In other words, $u$ \emph{covers} $v$. A \emph{block} is a connected subgraph with the same color vertices. Suppose $k$ is the number of blocks in the tree $\tree$. We define the block tree of $\tree$, denoted by $\mathcal{B}(\tree)$, as the tree with $k$ vertices, each corresponding to a block of $\tree$, and there is an edge between any two vertices if their corresponding blocks are neighbors in $\tree$. In other words, $\mathcal{B}(\tree)$ is obtained by contracting all blocks of $\tree$. Notice that each vertex of $\mathcal{B}(\tree)$ corresponds to a block of $\tree$, and vice versa. We refer to a block of $\tree$ as a \emph{leaf block} if its corresponding vertex in $\mathcal{B}(\tree)$ is a leaf. A block of $\tree$ that is not a leaf. The examples of blocks in a tree and the block tree are given in Figure \ref{1b}(A) and \ref{1b}(B), respectively. \emph{consistent spanning subset} (in short \css) is a variant of the consistent subset problem. A minimum consistent spanning subset (in short \mcss) is a minimum consistent subset with a restriction that every block of the graph must contain at least one vertex in the consistent subset. Note that not every block contains at least one vertex in the solution of \mcs, but \mcss takes at least one vertex from each block in the solution. For example, we present \mcs, \mcss and \mscs in Figure \ref{1a}(A), \ref{1a}(B) and \ref{1a}(C), respectively (vertices inside the circle are in the solution). The examples above demonstrate that \mcss and \mscs are distinct variations of \mcs. \mscs is an important version of \mcs because all neighbor vertices of a vertex $v$ in the solution set must have the same color as $v$.

\section{Previous Results and New Results}
The previous results on graphs and our new results are given below.
\subsection{Previous Results}
Banerjee et al.~\cite{Banerjee} demonstrated that the \mcs is \ with when parameterized by the size of the minimum consistent set, even with only two colors. Under basic complexity-theoretic assumptions for generic networks, a \fpt algorithm parameterized by $(c+\ell)$ is not feasible. This naturally raises the question of identifying the simplest graph classes for which the problem remains computationally intractable. Dey et al.~\cite{Sanjana} proposed a polynomial-time algorithm for \mcs on bi-colored paths, spiders, combs, etc. Also, Dey et al.~\cite{Anil} proposed a polynomial-time algorithm for bi-colored trees. Arimura et al.~\cite{Arunima} introduced a \xp algorithm parameterized by the number of colors $c$, with a running time of $\mathcal{O}(2^{4c}n^{2c+3})$. Recently, Manna et al. also showed that \mcs on trees is \np-complete in \cite{Bubai}. In the article \cite{bubai1}, Manna et al. demonstrated that fixed-parameter tractability has a faster running time for the same problem. Also, Manna presented an $(4\alpha +2)$ approximation algorithm of \mcs in interval graphs and \apx-hard in circle graphs in the paper \cite{manna2024minimum}. A recent paper \cite{biniaz2024minimum} implemented a polynomial time algorithm for \mcss in trees. We took some basic ideas from the articles \cite{biniaz2024minimum} and \cite{Sanjana} to present the following new results.

\subsection{New Results}
We demonstrate that \mscs is \np-hard in general graphs in section \ref{hardness}. We show a 2-approximation in trees in the same section. However, the 2-approximation algorithm is not important because we show $\mathcal{O}(n^4)$ polynomial-time algorithms for finding \mscs in trees in the section \ref{tree}. In section \ref{path}, we demonstrate faster polynomial-time algorithms in paths, spiders, and combs with running times $\mathcal{O}(n)$, $\mathcal{O}(n^2)$, $\mathcal{O}(n^3)$, respectively for finding \mscs.

\section{Preliminaries, \np-hardness in General Graphs and 2-Approximation Algorithm in Trees}\label{hardness}
Before discussing the algorithm for finding \mscs in trees, we are describing some basic properties as follows.
\subsection{Preliminaries}
\begin{observation}\label{observation1}
If all vertices of $\tree$ have the same color, i.e. $\tree$ is monochromatic, then every vertex of $\tree$ is a minimum strict consistent subset for $\tree$.
\end{observation}

\begin{lemma}\label{lemma2}
Every strict consistent subset has at least one vertex from each block of $\tree$.
\end{lemma}
\begin{proof}

Let $B_i$ be the block such that no vertex of $B_i$ appear in a strict consistent subset. As no vertex of $B_i$ is a strict selective vertex, then it is obvious that all the vertices of $B_i$ must be strictly covered by some of the vertices of a different block $B_j$ ($i\neq j$ and $C(B_i)= C(B_j)$). Let $v$ be a vertex in $B_i$, which is strictly covered by a vertex $u$ of $S_j$. We must have the shortest path (say $P$) between $v$ and $u$ such that $P$ must have at least vertex $w$ with a different colour than the colour of $S_i$; otherwise, $S_i$ and $S_j$ are identical, and our lemma is proved. 
$\\ \\$
As $C(w) \neq C(S_i)$, then $u$ is one of the nearest neighbors of $w$ and $w$ is not strictly covered because $C(u)\neq C(w)$. Now, if $w$ is contained in the strict consistent subset, then $v$ is not again strictly covered by $u$ because $dist(v,w) < dist(v,u)$ and $c(v) \neq c(w)$; a contraction. 

\end{proof}

As every block contains at least one vertex in the solution, so it is obvious that the vertices of $B$ must be covered, the vertices appear instead of taking $\alpha$ colors, we can put algorithms for only $2$ colors that must work for $\alpha$ colors. 

\begin{figure}[t]
\includegraphics[width=12cm]{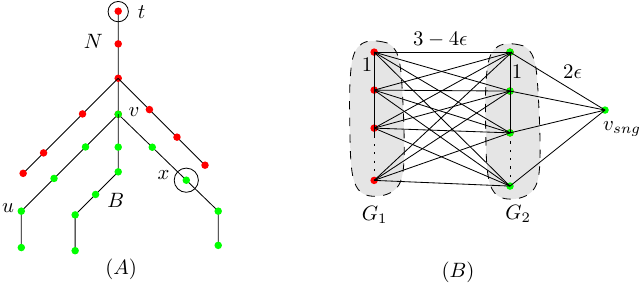}
\centering
\caption{(A) Illustration of Lemma \ref{lemma1}. (B) The reduced graph $G^{\prime}$.}\label{1c}
\end{figure}

\begin{lemma}\label{lemma1}
A minimum strict consistent subset of $\tree$ includes exactly one vertex from each leaf block.
\end{lemma}
\begin{proof}
We show this lemma by contradiction. Consider a minimum strict consistent subset $S^{\prime}$ with multiple vertices from leaf block $B$. Let $\neigh$ be the block next to $B$ (that is adjacent to $B$ in the block tree). Let $x$ be the nearest vertex of $S^{\prime}$ in $B$ to $\neigh$, as shown in Figure \ref{1c}(A). The set $S^{\prime\prime}$ is obtained by eliminating all vertices of B except $x$ from $S^{\prime}$. We argue that $S^{\prime}$ is a strict consistent subset of $\tree$, which contradicts that $S^{\prime}$ is minimum.

Now we show that $x$ is a vertex of $S^{\prime}$ that is closest to all vertex of $B$ to establish the above claim. Let $t$ be the nearest vertex of $S^{\prime}$ in $\neigh$ to $B$. The figure shows $v$ as the last vertex of $B$ on the path from $x$ to $t$. Since $S^{\prime}$ is a strict consistent subset, $\dist(v, x)< \dist(v, t)$. Now, examine any vertex in $u\in B$. The path from $u$ to $t$ travels through $v$. Thus, $\dist(u, x) < \dist(u, t)$, and our claim is proved.
\end{proof}

\subsection{\np-hardness of the \mscs in General Graphs}
We reduce our problem from the dominating set problem on graph $G$. We create a graph $G^{\prime} = G_1 \cup G_2 \cup \{v_{sng}\}$ from $G$, where $G_1$ and $G_2$ are copies of $G$ and $v_{sng}$ is a singleton vertex. Each edge of $G_1$ and $G_2$ has weight $1$. All vertices of $G_1$ are adjacent to all vertices of $G_2$ by edges. The weight of the edges between $G_1$ and $G_2$ is $3-4\epsilon$, where $0< \epsilon < \frac{1}{2}$. Additionally, all vertices of $G_2$ are directly connected by edges of weight $2\epsilon$ to $v_{sng}$ (Figure \ref{1c}(B)). Each vertex in $G_1$ is \emph{red} color, each vertex in $G_2$ is \emph{green} color, and $v_{sng}$ is also \emph{green} color. This completes the construction of our reduction.

\begin{lemma}\label{bubai1}
$G$ has a dominating set of size $k$ if and only if $G^{\prime}$ has a strict consistent subset of size $k + 1$.
\end{lemma}
\begin{proof}
Suppose that $G$ has a dominating set $S$ of size $k$. We will show that $G^{\prime}$ has a strict consistent subset of size $k+1$. Suppose that $S^{\prime}\leftarrow \emptyset$. For each vertex of $S$, we put the corresponding vertex in $G_1$ in $S^{\prime}$. Also, we put $v_{sng}$ in $S^{\prime}$. Therefore, $\lvert S^{\prime} \rvert =k+1$. We claim that $S^{\prime}$ is a strict consistent subset of $G^{\prime}$. For a red color vertex $v\in G_1$, if $v\notin S^{\prime}$, then $\dist(v,v_{sng})=3-4\epsilon +2\epsilon > 1$, but $v$ has a neighbor vertex in $G_1$ which is in $S^{\prime}$ of length $1$. Therefore, all the red-colored vertices are strictly covered. Also, for a green-colored vertex $u\in G_2$, $\dist (u,G_1)=3-4\epsilon > 2\epsilon =\dist (u,v_{sng})$. Therefore, all the red-colored vertices are strictly covered by $v_{sng}$.

Again, suppose that $S^{\prime}$ is a strict consistent subset of $G^{\prime}$ such that $\lvert S^{\prime} \rvert =k+1$. Therefore $v_{sng}$ must included in $S^{\prime}$; otherwise, size of $S^{\prime}$ must be greater than $k+1$. Therefore, all the vertices of $G_2$ are strictly covered. Hence we have $k$ vertices from $G_1$ in $S^{\prime}$. We take corresponding vertices of $G_1$ in $S^{\prime}$ into a set $S$. We claim that $S$ is the dominating set of $G$. If not then there exists a vertex $v\in S$ such that there is no neighbor of $v$ in $S$ and $v\notin S$. Therefore the corresponding vertex of $v$ in $G_1$ can not be covered strictly by $S^{\prime}$; a contradiction.
\end{proof}

\subsection{2-Approximation Algorithm of \mscs in Trees}
Consider the block tree $\mathcal{B}(\tree)$, as shown in Figure \ref{1b}. Let $S$ be an empty set.  Our technique adds the two vertices of $\tree$ that correspond to the endpoints of $e$ to $S$ for every edge $e \in \mathcal{B}(\tree)$. Clearly, $S$ is a feasible solution for the \mscs problem. We assert $S$ is a 2-approximate solution. Let $b$ represent the number of blocks in $\tree$, which is the same as the number of vertices in $\mathcal{B}(\tree)$. As $\mathcal{B}(\tree)$ is a tree, it has $b-1$ edges. Therefore, $\lvert S\rvert = 2b-2$. However, any optimal solution $OPT$ must contain at least one vertex from each block of $\tree$ (Lemma \ref{lemma2}). So, $\lvert OPT \rvert \geq b$. Therefore

$$\lvert S\rvert = 2b -2 \leq 2\lvert OPT\rvert -2.$$

This elementary analysis is the best for this algorithm. For example, if $\tree$ is a path with $b$ blocks, each with three vertices, our algorithm selects two vertices (except the two leaf blocks), and the optimal solution selects the middle vertex from each block. Each leaf block has exactly one vertex in $S$. Therefore, $\lvert OPT \rvert = b$ and $\lvert S \rvert =2b-2$, proving our assertion.

\section{Algorithm for Finding \mscs in Trees}\label{tree}
According to Lemma \ref{lemma1}, the \mscs problem can be more easier by taking a leaf block $B$ and requiring only one vertex to be in the solution. We try all the vertices of $B$ and report the best one.

\smallskip

We use nontrivial dynamic programming in our algorithm. We will first present the subproblems, and then we show how to solve subproblems recursively.

\subsection{Defining Subproblems in Trees}\label{subproblemForTree}
\begin{figure}[t]
\includegraphics[width=12cm]{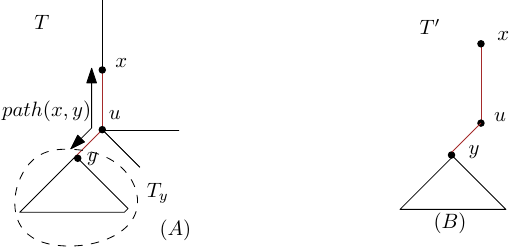}
\centering
\caption{(A) The tree $\tree$ , and (B) the tree $\tree ^{\prime}$.}\label{1d}
\end{figure}

Each subproblem is represented as $\tree (x, y)$, where $x$ and $y$ are two vertices of $\tree$. Consider $\patha (x,y)$ between $x$ and $y$ in $\tree$. Let $u\in \neigh (y)$ in $\patha (x,y)$ (possibly $x = u$). Removing the edge $(u, y)$ from $\tree$ yields two subtrees. Refer to Figure \ref{1d}(A) for the subtree containing $y$ known as $\tree_y$. Let $\tree^{\prime}$ be the union of $\paath(x,y)$ and $\tree_y$ (see to Figure \ref{1d}(B We define $\tree (x,y)$ as the \mscs problem on $\tree ^{\prime}$ with the following constraints:

\begin{itemize}
  \item the solution must include the vertex $x$, and
  \item $x$ must cover all vertices on $\paath(x,y)$, from $x$ to $u$.
\end{itemize}

It is clear from the above constraints that $x$ to $u$ should have the same color according to the definition of \mscs.

\subsection{Solving the Subproblems in Trees}

We denote the size of the (constrained) \mscs for $\tree (x,y)$ by $S(x, y)$. If $\tree (x, y)$ has no solution, then we set $S(x,y)=+\infty$. To solve $\tree (x, y)$ we proceed as follows

Observation \ref{observation1} states that if $\tree^{\prime}$ is monochromatic, then we return $x$ as a solution. In this situation, $S(x, y)=1$. Assume $\tree^{\prime}$ isn't monochromatic. Now, $x$ is the root of $\tree ^{\prime}$.

\begin{lemma}\label{lemma3}
Suppose $\tree(x, y)$ has a solution. Any solution of $\tree(x, y)$ has a vertex $z$ in the same or neighboring block of $x$ such that $x$ or $z$ covers all vertices on $\paath (x,z)$.
\end{lemma}
\begin{proof}
Any solution to $\tree(x, y)$ must have at least two vertices, as $\tree^{\prime}$ is multicolored. In particular, it should have a vertex from each neighboring block of $x$. Therefore, for such a vertex $z$, the lemma is satisfied by the vertex closest to $x$ in any solution of $\tree(x, y)$.
\end{proof}

\begin{figure}[t]
\includegraphics[width=12cm]{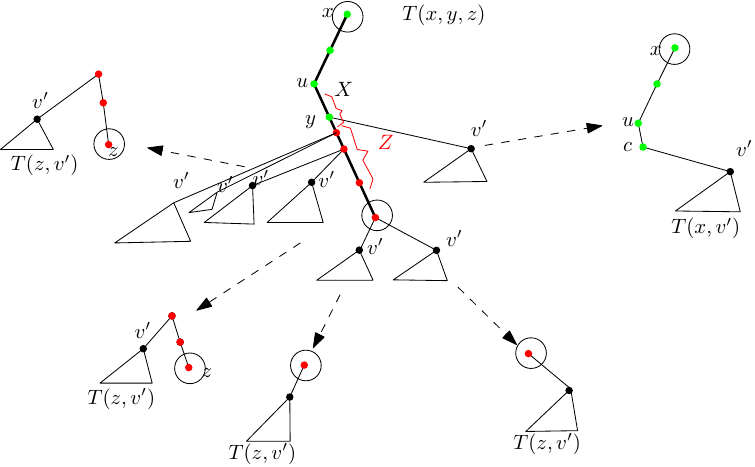}
\centering
\caption{Solving $\tree (x, y)$ recursively in terms of $\tree (x, v^{\prime})$ and $\tree (z, v ^{\prime} )$ where $z$ is a valid pair for $x$.}\label{1e}
\end{figure}

Let $z$ be any vertex of $\tree^{\prime}$ that satisfy the Lemma \ref{lemma3} (see Figure \ref{1e}). It also case that $z = y$ may happen. Also, $z$ could be in $x$'s block or its neighbor. If a vertex $z$ does not exist, $\tree(x, y)$ has no solution, hence $S(x, y) = +\infty$.

\begin{figure}[t]
\includegraphics[width=12cm]{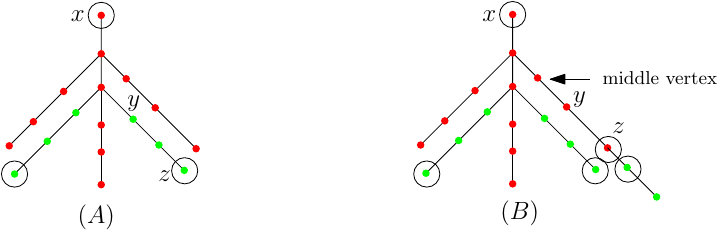}
\centering
\caption{(A) The tree $\tree$ , and (B) the tree $\tree ^{\prime}$.}\label{1k}
\end{figure}

Lemma \ref{lemma3} states that $x$ or $z$ covers all vertices of $\paath(x,z)$. The definition of $\tree(x, y)$ requires that $u$ be covered by $x$, thus $\dist(u, x) < \dist(u, z)$ and $\dist(z, x) > 2 \cdot \dist(u, x)$. Furthermore, if $x$ and $z$ have different colors, $\paath(x,z)$ has an even number of vertices (see Figure \ref{1k}(A)); otherwise, it does not satisfy the definition of \mscs. Assume $\paath(x,z)$ has $2k$ vertices. The first $k$ vertices must be covered by $x$, and the second $k$ by $z$. If $x$ and $z$ have different colors, then it does not matter whether length of $\paath(x,z)$ is even or odd (see Figure \ref{1k}(B)).

Now we see that both $x$ and $z$ must be present in the solution, and they must cover all vertices on $\paath (x,z)$. We represent this more constrained version of $\tree(x, y)$ as $\tree (x, y, z)$ (we do not call this a subproblem for a reason to be determined later). In other words, $\tree(x, y, z)$ represents the \mscs problem on $\tree ^{\prime}$ with the following constraints:

\begin{itemize}
  \item $z$ is in the same block as $x$ or in a neighboring block of $x$.
  \item The solution must include both $x$ and $z$.
  \item $\dist(z, x) > 2 \cdot \dist(u, x)$.
  \item $\paath (x,z)$ has $2k$ where the first $k$ vertices must have the same color as $x$ and the second $k$ vertices must have the same color as $z$ when $C(x)\neq C(z)$.
  \item If $C(x)=C(z)$, then the length of the path $\paath (x,z)$ can be even or odd. If length of the path $\paath (x,z)$ is odd then the middle vertex is covered by anyone of $x$ and $z$ (see Figure \ref{1k}(B)).
\end{itemize}

A \emph{valid pair} for $x$ is defined as any vertex $z$ that satisfies the above constraints. Now we demonstrate how to solve $\tree(x, y, z)$. The size of the solution for $\tree(x, y, z)$ is denoted as $S(x, y, z)$. Let $X$ be the set of all vertices on $\paath(y,z)$ that are closer to $x$ than to $z$, as shown in Figure \ref{1e}. Let $Z$ denote the set of all vertices on $\paath (y,z)$ that are closer to $z$ than to $x$.

To solve $\tree(x, y, z)$, we define two sets $\mathcal{X}$ and $\mathcal{Z}$ as shown below. For each vertex $v$ in $X$, we add to $\mathcal{X}$ those children of $v$ who are not on the path $\paath(y,z)$. For each vertex $v$ in $Z$, we add to $\mathcal{Z}$ all children of $v$ who are not on the path $\paath(y,z)$. The solution of $\tree(x, y, z)$ is found by taking the union of $\{x, z\}$ with the solutions of $\tree(x, v^{\prime})$ for all $v^{\prime}\in \mathcal{X}$ and the solutions of $\tree(z, v^{\prime})$ for all $v^{\prime}\in \mathcal{Z}$. It might be a case that not all choices of $z$ may result in a valid solution. In such a case, $\tree(z, v^{\prime})$ will be infeasible, and $S(z, v^{\prime}) = +\infty$.

For two vertices $v^{\prime}_1, v^{\prime}_2 \in \mathcal{Z}$, the solutions of $\tree(z, v^{\prime}_1)$ and $\tree(z, v^{\prime}_2)$ may affect each other. However, this cannot happen because by the definition of $\tree (.,.)$ all vertices on paths from $z$ to the parents of $v^{\prime}_1$ and $v^{\prime}_2$ must be covered by $z$, and thus any vertex in the solutions of $\tree (z, v^{\prime}_1)$ and $\tree(z, v^{\prime}_2)$ lies in the same level as $z$ or in a lower level (i.e. higher depth) in $\tree ^{\prime}$. The same argument applies to vertices in $\mathcal{X}$. In the final solution, no vertex has an equal distance to multiple vertices of different colors due to the definition of \mscs. Therefore, 

$$ S(x,y,z)= 2+ \sum_{v^{\prime}\in \mathcal{X}}S(x,v^{\prime})+\sum_{v^{\prime}\in \mathcal{Z}}S(z,v^{\prime})-\lvert \mathcal{X}\rvert - \lvert \mathcal{Z}\rvert,$$

where the subtractive terms $\lvert \mathcal{X}\rvert$ and $\lvert \mathcal{Z}\rvert$ come from the fact that $x$ and $z$ are counted in each $S(x, v^{\prime})$ and $S(z, v^{\prime})$. Thus, we are able to solve $\tree(x, y, z)$ in terms of $\tree(x, v^{\prime})$ and $\tree(z, v^{\prime})$ which
are smaller instances of $\tree(x, y)$.

Now we return to the original problem, $\tree(x, y)$. If we knew $z$, we can solve $T(x, y, z)$ as well as $T(x, y)$. However, we do not know $z$ in advance. As a result, we consider all valid $z$ combinations for $x$ and select the one that gives us the smallest solution. Let $P$ be the collection of all valid pairs for $x$. Then 

$$S(x,y)=\min \{S(x,y,z) \text{ } |\text{ } z\in P\}.$$

If $P$ is the empty set, which means $x$ has no valid pairs, we set $S(x, y) = +\infty$. This concludes our solution to the subproblem $\tree(x, y)$. The problem $T(x, y, z)$ was constructed to simplify the presentation of our recursive solution.

\subsection{Solving the Original Problem in Trees}

This section demonstrates how to solve the original \mscs problem for the tree $\tree$. Let $S$ denote an optimal solution to this problem. If $\tree$ is monochromatic, we return any vertex as a solution. Assume $\tree$ has more than one color. Also, assume that $L$ is a leaf block of $\tree$, and $\neigh$ is its only neighboring block. Let $y$ be the vertex of $\neigh$ adjacent to a vertex of $L$. By Lemma \ref{lemma1}, any optimal solution has exactly one vertex, say $x$, from L. Thus, $x$ is the closest vertex of $S$ to all
vertices in $L$. Hence the original problem is an instance of the $\tree(x, y)$ problem that was introduced earlier. Since we do not know $x$, we try all vertices of $L$. Therefore,

$$\lvert S\rvert=\min \{S(x,y) \text{ } |\text{ } x\in L\}.$$

The correctness of our algorithm is implied by Lemma \ref{lemma1}, the fact that in any optimal solution, all the vertices of $L$ are closer to $x$ than to any other vertex of $S$, and from the correctness of our solution for $\tree(x, y)$.

\subsection{Running Time Analysis}
In this section we analyze the running time of our algorithm for the \mscs problem on a tree $\tree$ with $n$ vertices. The algorithm follows a top-down dynamic programming approach and consists of subproblems of the form $\tree(x, y)$ where $x$ and $y$ are two vertices of $\tree$. In total we have $\mathcal{O}(n^2)$ subproblems of this form. To solve each subproblem we examine $\mathcal{O}(n)$ valid pairs for $x$. For each valid pair $z$ of $x$ we look up to solutions of $\mathcal{O}(n)$ smaller subproblems, namely $\tree(x, v^{\prime})$ and $\tree(z, v^{\prime})$. Thus each subproblem $\tree(x, y)$ can be solved in $\mathcal{O}(n^2)$ time. Therefore the total running time of the algorithm is $\mathcal{O}(n^4)$.



\subsection{Extension to Weighted Trees}

Our algorithm can simply be extended to solve the \mscs problem on weighted trees within the
same time complexity. The only parts that need to be adjusted are the definitions of distances and valid pairs. The distance $\dist(x, y)$ between two vertices $x$ and $y$ is now the the shortest distance in the weighted tree. The existence of a valid pair $z$ for a (Lemma \ref{lemma1}) can be shown by taking a vertex in the solution whose weighted distance to $x$ is minimum. The validity of $z$ is now verified by similar constraints, except for the last constraint, which we adjust as follows:
\begin{enumerate}
\item All vertices on path $\paath (x,z)$ that are closer to $x$ (in terms of weighted distance) must have the same color as $x$ and all vertices that are closer to $z$ must have the same color as $z$.
\end{enumerate}

\section{Algorithm for Finding \mscs in Paths, Spiders, Combs}\label{path}
The definitions of the paths, spiders and combs must be found in \cite{diestel2012graph} and \cite{Sanjana}. Therefore, the algorithms in paths, spiders, and combs are as follows:
\subsection{Algorithm for Finding \mscs in Paths}\label{pathhh}
\begin{figure}[t]
\includegraphics[width=12cm]{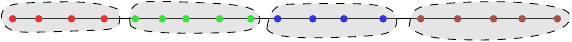}
\centering
\caption{Blocks of a path}\label{1g}
\end{figure}

In a path graph $G$, the vertices in $V(G)$ are listed in the order $p_1, p_2,\dots,p_n$. Each pair of consecutive vertices define an edge of the graph, i.e., $E(G) = \{(p_i, p_{i+1}), i = 1, 2,\dots,n -1\}$. Each vertex has degree two except the two terminal vertices of the path that have degree one. In a $\alpha$-chromatic path graph, each vertex is assigned with one color in $\{c_1, \dots, c_{\alpha}\}$. The blocks of the path are shown in Figure \ref{1g}. In a block of length 1, the vertex itself is always selected.
\begin{lemma}\label{bubai2}
In the minimum strict consistent subset of a path graph, each block will
have at least one and at most two vertices in $S$. Moreover, exactly one vertex will be sufficient from the first and the last block.
\end{lemma}
\begin{proof}
The presence of at least one element from each block in $S$ is obvious by using Lemma \ref{lemma2}. It is also obvious that if more than two vertices from a block $B$ are chosen in $S$ then all the chosen vertices in that block excepting the first and last one can be dropped without violating the strict consistency, and thereby reducing the size of $S$. By the same argument if more than one vertex from the first (resp. last) block is chosen in $S$ then all those chosen vertices excepting the rightmost in the first block (resp. leftmost in the last block) can also be dropped without violating the consistency.
\end{proof}

\smallskip

\begin{figure}[t]
\includegraphics[width=7cm]{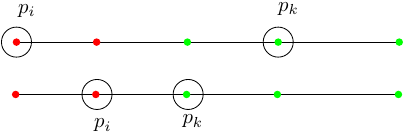}
\centering
\caption{Valid pair $(p_i, p_k)$}\label{1h}
\end{figure}

\noindent \underline{\textbf{Algorithm:}} Consider a pair of adjacent blocks $B_j$ and $B_{j+1}$. Assume, without
loss of generality, $\lvert B_j \rvert \leq \lvert B_{j+1}\rvert$. For each member $p_i \in B_j$, there exists at most one vertex, say $p_k\in B_{j+1}$, such that if $p_i$ is included in $S$ then $p_k$ of $B_{j+1}$ must be included in $S$ to satisfy the strict consistency property of the boundary vertices of $B_j$ and $B_{j+1}$ that are adjacent to each other (see Figure \ref{1h}). Thus, $(pi, p_{k})$ forms a \emph{valid-pair}. Now, we define the overlay graph $H$ as follows. The vertices of $H$ are the vertices of $G$, and $k$ dummy vertices $D = \{d_1,\dots,d_k\}$, where $k$ is the number
of blocks. Therefore, $V(H)=V(G)\cup D$. The edges in the set $E(H)$ are of two types. For each valid-pair, we add a directed type-1 edge in $V(H)$. For each vertex $p_i$ in a block $B_j$ we add two directed type-2 edges $(p_i, d_j )$ and $(d_j , p_i)$ in $V(H)$ . The weight of each type-1 edge is $0$. The type-2 edges incident to $D\backslash \{d_1, d_k\}$ have weight $1$. Each type-2 edge incident
to ${d_1, d_k}$ has weight $0$. For the complete demonstration of the graph $H$. A forward $s-t$ path is a path from $s$ to $t$ where the indices of the $p_i$ vertices appear in increasing order. Now we find the shortest forward $s-t$ path with $s = d_1$ and $t = d_k$ in the graph $H$, and remove the $d_j$’s to obtain \mscs of the original path graph $G$.

\begin{theorem}\label{bubai3}
The shortest $s-t$ path of the overlay graph $H$ gives the minimum strict consistent subset of the path G, and it executes in $\mathcal{O}(n)$ time.
\end{theorem}
\begin{proof}
We first prove that any forward $s-t$ path of the graph $H$ constructed from the given path graph $G$ gives a strict consistent subset of the graph $G$. Observe that at least one vertex from each block is present in the $s-t$ path of the graph H. The reason is that the edges corresponding to the valid pairs are defined only between adjacent blocks in forward direction, and each dummy vertex (say $d_i$) is bidirectionally connected with only the vertices of one block in the graph $H$. Each edge $(v_i,v_j)$ of $H$ between two consecutive blocks in $G$ justify the strict consistency of the vertices $\{v_i,v_{i+1},\dots,v_j\}$ of the path graph $G$. As all the vertices in any path of $H$ between a pair of vertices $v_i$, $v_j$ in the same block are of same color, choice of those vertices on the path in the strict consistent subset does not destroy the consistency property of the not chosen vertices between $v_i$ and $v_j$. Thus, any $s-t$ path with $s = d_1$ and $t = d_r$ in $H$ gives a strict consistent subset of the vertices in the path graph $G$.

Now, we will consider the nature of the minimum $s-t$ path in $H$. Each forward move in the minimum $s-t$ path from a vertex $v_i \in B_j$ either reaches a vertex $v_{\ell} \in B_{j+1}$ or to a vertex $v_m \in R_j$ through the dummy vertex $d_j$ where $i < m$. If no dummy vertex in $\{d_2,\dots,d_{k-1}\}$ is visited in the $s-t$ path then exactly one vertex is present from each block in the obtained path. However, the presence of every dummy vertex $d_i$ in the shortest $s-t$ path implies that two vertices of the corresponding block is present in the consistent subset obtained by that $s-t$ path. 

Also, the construction of the graph suggests that if a subset of vertices in $V(H) \cup D$ do not form a path, then they cannot form a consistent subset. The minimality in the size of the consistent subset is justified from the choice of the shortest $s-t$ path.

The number of type-1 edges in the graph $H$ is at most $n$ since each vertex in a block $G$ can participate in at most $1$ valid pair in its succeeding block. The number of type-2 edges is $2n$ since each vertex in $G$ is bidirectionally connected with the dummy vertex of its corresponding block. Thus we have $\mathcal{O}(n)$ edges in $H$. In the special case of integer weights and directed connected graphs, Dijkstra’s algorithm for shortest path executes in $\mathcal{O}(|E(G)|)$ time \cite{10.1145/316542.316548}. Thus, the time complexity follows.
\end{proof}

\smallskip

\noindent\textbf{Remark:} A minor tailoring of the same algorithm works for a $\alpha$-chromatic cycle
graph $G$, where the vertices in $V(G)$ are connected in a closed chain.

\subsection{Algorithm for Finding \mscs in Spiders}
Let $B$ denote the block containing the center of the spider graph. Any solution must contain at least one vertex from $B$ according to Lemma \ref{lemma2}, among which one is closest to the center. We examine each vertex $b \in B$ as a vertex in the solution that is closest to the center. Then, we break the spider into some paths, each of which is defined by two endpoints $b$ and the end vertex of the leg of the spider. That is, for each leg of the spider, we take the end vertex of the leg and the vertex $b$, which forms a path. We solve the \mscs problem on each path independently (with the constraint that $b$ must be in the solution) and then take the union of the solutions as a solution for the \mscs problem in the spider. The total running time is $\lvert B\rvert \cdot \mathcal{O}(n) = \mathcal{O}(n^2)$.

\subsection{Algorithm for Finding \mscs in Combs}
We use our algorithm of previous section (for general trees) and show how to save a linear factor from the running time.

Recall from subsection \ref{subproblemForTree} that we solve $\tree (x, y, z)$ in linear time by looking up to solutions of $\mathcal{O}(n)$ smaller subproblems of the form $\tree (x, v^{\prime})$ and $\tree(z, v^{\prime})$. We are going to solve this in constant time with the help of pre-processing. We take advantage of the fact that each vertex $v$ on the path $\paath (x,z)$ has at most one child that can be $v^{\prime}$ (because $\tree$ is a comb) and the fact that each subproblem $T(x, v^{\prime})$ is an \mscs instance on a path (again because $\tree$ is a comb).

Let $s_1,\dots, s_k$ be the vertices of the skeleton in this order. For each $i \in \{1,\dots, k\}$ let $l_i$ be the leaf of $\tree$ in the dangling path at $s_i$. We define two $n\times k$ matrices $\mathcal{P}$ and $\mathcal{Q}$ where the rows represent the $n$ verties of $\tree$ and the columns represent the $k$ vertices of the skeleton. For each vertex $x \in \tree$ and each $i \in \{1,\dots, k\}$ let $l_i$ the entry $\mathcal{P}[x][i]$ stores the size of the solution for the \mscs problem on the path from $x$ to $l_i$ with the constraint that $a$ is in the solution and all vertices from $a$ to $s_i$ are covered by $x$ and this can be computed in linear time using path algorithm. Each entry $\mathcal{Q}[x][i]$ stores the sum of $\mathcal{P}[a][j]$ for all indices $j$ of skeleton vertices that lie on the path $\paath(x,s_i)$. Thus $\mathcal{P}$ and $\mathcal{Q}$ can be computed in $\mathcal{O}(kn^2)$ and $\mathcal{O}(k^2n)$ time, respectively, in a pre-processing step.

Let $m$ be the index such that all vertices on the path $\paath(x,p_m)$ have the same color as $x$ and all vertices on the path $\paath (p_{m+1},z)$ have the same color as $z$ (this corresponds to definitions of the sets $A$ and $Z$ in the original algorithm). Then $\mathcal{Q}[x][m]=\sum_{v^{\prime}\in \mathcal{A}}S(x,v^{\prime})$ and $\mathcal{Q}[z][m+1]=\sum_{v^{\prime}\in \mathcal{Z}}S(z,v^{\prime})$. Since $\mathcal{Q}[x][m]$ and $\mathcal{Q}[z][m+1]$ are already computed, we can solve $\tree(x, y, z)$ in constant time. Therefore, the total running time of the algorithm is $\mathcal{O}(n^3)$.

\section{Concluding Remarks}
The problem \mscs is a variant of the \mcs. Hence, the algorithms for \mscs problems in other graph classes are open, especially for planar graphs. Furthermore, as we demonstrate that \mscs is \np-hard in general graphs, the approximation algorithm, \fpt, \apx-hard, and \ptas, are also open for general graphs.

\bibliographystyle{unsrt} 
\bibliography{references}  

\begin{thebibliography}{10}

\bibitem{Hart}
Peter~E. Hart.
\newblock The condensed nearest neighbor rule (corresp.).
\newblock {\em {IEEE} Trans. Inf. Theory}, 14(3):515--516, 1968.

\bibitem{Wilfong}
Gordon~T. Wilfong.
\newblock Nearest neighbor problems.
\newblock In Robert L.~Scot Drysdale, editor, {\em Proceedings of the Seventh Annual Symposium on Computational Geometry, North Conway, NH, USA, , June 10-12, 1991}, pages 224--233. {ACM}, 1991.

\bibitem{Bodhayan}
Kamyar Khodamoradi, Ramesh Krishnamurti, and Bodhayan Roy.
\newblock Consistent subset problem with two labels.
\newblock In B.~S. Panda and Partha~P. Goswami, editors, {\em Algorithms and Discrete Applied Mathematics - 4th International Conference, {CALDAM} 2018, Guwahati, India, February 15-17, 2018, Proceedings}, volume 10743 of {\em Lecture Notes in Computer Science}, pages 131--142. Springer, 2018.

\bibitem{Chitnis}
Rajesh Chitnis.
\newblock Refined lower bounds for nearest neighbor condensation.
\newblock In Sanjoy Dasgupta and Nika Haghtalab, editors, {\em Proceedings of The 33rd International Conference on Algorithmic Learning Theory}, volume 167 of {\em Proceedings of Machine Learning Research}, pages 262--281. PMLR, 29 Mar--01 Apr 2022.

\bibitem{Biniaz}
Ahmad Biniaz, Sergio Cabello, Paz Carmi, Jean{-}Lou~De Carufel, Anil Maheshwari, Saeed Mehrabi, and Michiel H.~M. Smid.
\newblock On the minimum consistent subset problem.
\newblock In Zachary Friggstad, J{\"{o}}rg{-}R{\"{u}}diger Sack, and Mohammad~R. Salavatipour, editors, {\em Algorithms and Data Structures - 16th International Symposium, {WADS} 2019, Edmonton, AB, Canada, August 5-7, 2019, Proceedings}, volume 11646 of {\em Lecture Notes in Computer Science}, pages 155--167. Springer, 2019.

\bibitem{diestel2012graph}
Reinhard Diestel.
\newblock Graph theory, volume 173 of.
\newblock {\em Graduate texts in mathematics}, page~7, 2012.

\bibitem{Banerjee}
Sandip Banerjee, Sujoy Bhore, and Rajesh Chitnis.
\newblock Algorithms and hardness results for nearest neighbor problems in bicolored point sets.
\newblock In Michael~A. Bender, Martin Farach{-}Colton, and Miguel~A. Mosteiro, editors, {\em {LATIN} 2018: Theoretical Informatics - 13th Latin American Symposium, Buenos Aires, Argentina, April 16-19, 2018, Proceedings}, volume 10807 of {\em Lecture Notes in Computer Science}, pages 80--93. Springer, 2018.

\bibitem{Sanjana}
Sanjana Dey, Anil Maheshwari, and Subhas~C. Nandy.
\newblock Minimum consistent subset of simple graph classes.
\newblock In Apurva Mudgal and C.~R. Subramanian, editors, {\em Algorithms and Discrete Applied Mathematics - 7th International Conference, {CALDAM} 2021, Rupnagar, India, February 11-13, 2021, Proceedings}, volume 12601 of {\em Lecture Notes in Computer Science}, pages 471--484. Springer, 2021.

\bibitem{Anil}
Sanjana Dey, Anil Maheshwari, and Subhas~C. Nandy.
\newblock Minimum consistent subset problem for trees.
\newblock In Evripidis Bampis and Aris Pagourtzis, editors, {\em Fundamentals of Computation Theory - 23rd International Symposium, {FCT} 2021, Athens, Greece, September 12-15, 2021, Proceedings}, volume 12867 of {\em Lecture Notes in Computer Science}, pages 204--216. Springer, 2021.

\bibitem{Arunima}
Hiroki Arimura, Tatsuya Gima, Yasuaki Kobayashi, Hiroomi Nochide, and Yota Otachi.
\newblock Minimum consistent subset for trees revisited.
\newblock {\em CoRR}, abs/2305.07259, 2023.

\bibitem{Bubai}
Bubai Manna and Bodhayan Roy.
\newblock Some results on minimum consistent subsets of trees.
\newblock {\em CoRR}, abs/2303.02337, 2023.

\bibitem{bubai1}
Aritra Banik, Sayani Das, Anil Maheshwari, Bubai Manna, Subhas~C Nandy, Bodhayan Roy, Sasanka Roy, Abhishek Sahu, et~al.
\newblock Minimum consistent subset in trees and interval graphs.
\newblock {\em arXiv preprint arXiv:2404.15487}, 2024.

\bibitem{manna2024minimum}
Bubai Manna.
\newblock Minimum consistent subset in interval graphs and circle graphs.
\newblock {\em arXiv preprint arXiv:2405.14493}, 2024.

\bibitem{biniaz2024minimum}
Ahmad Biniaz and Parham Khamsepour.
\newblock The minimum consistent spanning subset problem on trees.
\newblock {\em Journal of Graph Algorithms and Applications}, 28(1):81--93, 2024.

\bibitem{10.1145/316542.316548}
Mikkel Thorup.
\newblock Undirected single-source shortest paths with positive integer weights in linear time.
\newblock {\em J. ACM}, 46(3):362–394, may 1999.

\end{thebibliography}
\end{document}